# Imaging Ferroelectrics: Charge Gradient Microscopy (CGM) versus Potential Gradient Microscopy (PGM)

Jesi R. Maguire[1], Hamza Waseem[1], Raymond G. P. McQuaid[1], Amit Kumar[1], J. Marty Gregg[1] and Charlotte Cochard[2]

[1] *School of Mathematics and Physics, Queen's University Belfast, Belfast BT7 1NN, United Kingdom*

[2] *School of Science and Engineering, University of Dundee, Nethergate, Dundee DD1 4HN, United Kingdom*

## Abstract

In 2014, Charge Gradient Microscopy (CGM) was first reported as a new scanning probe imaging mode, particularly well-suited for the characterisation of ferroelectrics. The implementation of the technique is straightforward; it involves monitoring currents that spontaneously develop between a passive conducting atomic force microscopy tip and Earth, as the tip is scanned across the specimen surface. However, details on the fundamental origin of contrast and what images mean, in terms of associated ferroelectric microstructures, are not yet fully understood. Here, by comparing information from CGM and Kelvin Probe Force Microscopy (KPFM), obtained from the same sets of ferroelectric domains (in both lithium niobate and barium titanate), we show that CGM reasonably reflects the spatial derivative of the measured surface potential. This is conceptually different from measuring local gradients in the surface bound-charge density or in any associated screening charges: after all, we see clear CGM signals, even when polarisation is entirely in-plane. We therefore suggest that CGM in ferroelectrics might be more accurately called Potential Gradient Microscopy (PGM). Intriguingly, in all cases examined, the measured surface potential (determined both through KPFM and by integrating the CGM signal) is of the opposite sign to that intuitively expected for a completely clean ferroelectric in vacuum. This is commonly observed and presumed due to a charge accumulation on the ferroelectric surface which is not easily removed.

# 1. Introduction

Historically, technological developments in microscopy have been extremely important for underpinning new insights into the nature and behaviour of ferroelectric materials. Over the last few decades, this has perhaps been most strikingly illustrated by the way in which aberration-corrected transmission electron microscopy (TEM) has facilitated picometre resolution images of local dipole vectors. Flux-closure [1], vortex [2] and even skyrmion structures [3] have been stunningly revealed. Combining TEM with focused ion beam (FIB) microscope patterning has also allowed the response of ferroelectric domain patterns to nanoscale morphology to be firmly established [4,5]. The importance of advances in scanning probe microscopy (SPM) should also be fully recognised: piezoresponse force microscopy (PFM) and its precursors have revolutionised the way in which ferroelectric domain microstructures can be mapped [6-9] and their dynamics studied [10-14]; conducting atomic force microscopy (cAFM) has been *the* breakthrough tool for unequivocally showing that domain walls can have profoundly different transport characteristics from the domains that they surround [15-20]; in addition, scanning nitrogen-vacancy magnetometry has exquisitely revealed spin cycloid microstructures in multiferroics [21] and has even been suggested as a future technique for examining local transient magnetic fields, expected during ferroelectric domain wall motion [22].

One of the most recently discovered SPM imaging modes, relevant for ferroelectric characterisation (first reported by Hong *et al.* [23] in 2014), is Charge Gradient Microscopy (CGM). Practically, CGM involves scanning a conducting atomic force microscopy tip (usually solid platinum), in contact mode at a relatively high set-point, so that a firm connection between tip and sample is maintained. Imaging is passive, in the sense that no deliberate external electrical signal is supplied. Instead, any currents that spontaneously develop between the tip and Earth are simply measured and recorded as a function of tip position. A key observation is that, while the magnitude of local current spikes may be increased by increasing the scan speed during imaging, the total integrated charge (flowing to or from Earth) is independent of the rate of tip motion relative to the sample surface. Contrast is therefore associated with the way in which the local tip-sample electrostatic interactions lead to changes in the charge state at the tip; such changes can be developed quickly (generating larger currents) or slowly (generating smaller currents), depending on the scanning rate used. As noted in Hong *et al.*'s original paper, CGM gives strong signals from c+/c- 180º domain walls on z-cut periodically poled lithium niobate surfaces, where obvious spatial gradients in bound charge densities (and in any associated screening charge densities) should be expected [23,24]. Hence the name given to the imaging mode. The exact nature of the interactions between the tip and the ferroelectric surface that give rise to changes in the tip charge state, relative to Earth, are still a matter of some debate and the role of charged surface adsorbates seems to be particularly unclear [23, 25].

In this paper, we do not make comment on the detail of the interaction between the tip and the screening charge. Instead, we show that the integration of the CGM current, as a function of tip position, maps well to the experimentally determined spatial variation in the electrostatic potential, measured directly using Kelvin Probe Force Microscopy (KPFM), for a number of different domain configurations: c+/c- 180º domains in periodically poled z-cut lithium niobate (with polarisation vectors perpendicular to the imaged surface), head-to-head 180º domains in x-cut lithium niobate (with polarisation vectors parallel to the imaged surface) and a-c domains in {100} polished barium titanate (with polarisation alternately oriented parallel and perpendicular to the imaged surface). Results suggest that, while the CGM technique undoubtedly monitors changes in the charge state of the conducting tip, this charge state appears to be more directly related to spatial variations in the surface potential of the ferroelectric than to spatial variations of the surface bound charge density (or the density of any associated screening charges) occurring directly under the tip. In this sense CGM should be viewed as effectively imaging potential gradients, rather than charge gradients, on the ferroelectric surface.

## 2. Results and Discussion

Initially, we established Charge Gradient Microscopy (CGM) images, similar to those obtained by Hong *et al.* [23], on z-cut periodically poled lithium niobate single crystals. Confirming their original observations [23], the correlation between the spatial occurrence of currents and the domain walls separating 180º domains (measured by PFM) was found to be obvious (figure 1, panels a, c and e). However, CGM contrast on engineered head-to-head domain walls in x-cut lithium niobate, where the polarisation is almost entirely parallel to the imaged surface, was seen to be dramatically different (figure 1, panels b, d and f). Instead of current peaks at the domain walls, where the spatial orientation of polarisation changes, here the dominant response was that the current was almost constant within each domain but changed sign as the polarisation orientation reversed (seen clearly in the colour contrast in figure 1 panel d). In detail, a peak at the wall was also present (figure 1 panel f); we attribute this to a slight miscut angle, introduced by our inexact surface polishing, which created a small component of polarisation out of the plane, which reverses sense across the domain wall.

These observations on lithium niobate (both z and x-cut) mirror those seen previously by Guy *et al.* [26], where the link between CGM current and the change, or gradient, in surface potential (rather than the change in uncompensated bound charge at the surface) was first suspected. In Guy *et al.*, local potential variations were, however, solely informed by finite element numerical simulations (Comsol multiphysics models). Here, we generate more meaningful insight by comparing CGM information with experimentally determined surface potentials, measured directly using Kelvin Probe Force Microscopy (KPFM).

KPFM imaging of ferroelectric surfaces in ambient conditions is a challenge, as surface adsorbates, which screen the local potential, accumulate over a timescale of between seconds and minutes. The KPFM images shown in figure 2 (panels c and d) were obtained directly after the domains had been imaged in contact mode (panels a and b) to clean the surface, by physically scraping most of such adsorbates away. Correlations between the potential contrast and the domain contrast are obvious. More subtle insight is, however, revealed by turning the slow-scan axis, in both CGM and KPFM imaging modes, off and accumulating signal from repeated scans along the same line-trace, to increase the signal-to-noise ratio in the data. In figure 3, aggregated KPFM line scans, taken perpendicular to the surface-trace of domain walls, for both c+/c- domains in z-cut lithium niobate (figure 3a) and in-plane head-to-head domains in x-cut lithium niobate (figure 3b), are presented. Rather than comparing these data to the raw CGM traces, they are instead compared to the spatial integrals of the CGM currents (figure 3, panels c and d). For c+/c- domains, both the KPFM surface potential and the integrated CGM currents show peaks above domains with downwards orientations of polarisation and troughs above those with upwards oriented polarisation (figure 3 panels a and c). Equally, for the head-to-head in-plane domains, KPFM reveals a strong "V"-shaped function, with the potential being at a minimum close to the wall (figure 3b). The same kind of "V" is seen for the spatial integration of the CGM currents, taken from the same microstructural region (figure 3d). While the match is not perfect, similarity in the form of these functions is clearly evident. We are therefore left with the following observation:

$$V \equiv \int_{x_1}^{x_2} \frac{dV}{dx}dx = k \int_{x_1}^{x_2} I_{\text{CGM}}dx \qquad \textbf{equation 1}$$

where $x$ is the position along the line-scan, $V$ is the local surface potential measured by KPFM in a small region (pixel) bounded by points $x_1$ and $x_2$, and $I_{\text{CGM}}$ is the locally measured CGM current; *k* is a proportionality constant dependent on the microscope setup. The implication from equation 1 is that the CGM current is equivalent to the gradient in the surface potential along the scanning direction:

$$I_{\text{CGM}} \propto \frac{dV}{dx}$$

To test this notion further, CGM and KPFM information was obtained for another set of domains, with both in-plane and out-of-plane polarisation present: a-c stripe domains in {100}$_{pseudocubic}$ polished single crystal BaTiO$_3$. The typical surface domain microstructure can be seen in the PFM maps presented in figure 4a. While the CGM maps (figure 4b) show stripe-like features which clearly correlate with the PFM microstructure, there is unexpected complexity apparent in the contrast. This is explicitly demonstrated in the multiple pass CGM line-scan (slow-scan axis turned off) shown in figure 4c, taken perpendicular to the surface trace of the domain walls: multiple current maxima and minima can be seen within each a-c domain period. When the CGM current response is averaged over a number of a-c periods (see supplementary information) and then spatially integrated, an "idealised" version of what the surface potential might look like can be inferred, assuming that equation 1 and its implications are true (figure 4d).

The associated directly measured potential (using KPFM) for the same microstructural region, is shown in figure 4e. These data have also been averaged for a and c-domains separately and the resultant functions "stitched" together (see supplementary information). At first sight, the potential profiles implied by the integrated CGM and directly measured KPFM look rather different in this case. However, by close examination of the form of the functions, within the same microstructural subregions, similarities can be identified: the regions contained within the yellow boxes in both figure panels (part of the a-domain) show distinct sharp potential minima; within the green boxes (the c-domain), two maxima and a minimum are seen in both plots; and the purple boxes (another part of the a-domain) illustrate transition regions, in which the potential generally increases as the a-domain is traversed. It therefore still seems possible that the integrated CGM and the KPFM both represent maps of the same ferroelectric surface potential (albeit imperfectly).

An aspect of the experimental work that should also be discussed concerns the signs of the CGM current, inferred surface potential determined by spatial integration of the CGM, and the directly measured surface potential. In figure 5, we have used finite element simulation (Comsol Mutiphyiscs) to calculate the surface potential of lithium niobate for both c+/c- domains in the z-cut crystal and for the in-plane head-to-head domain wall region in the x-cut crystal, in the absence of any screening. In the former case, domains with polarisation pointing out-of-the-plane have positive surface potentials, while those pointing into the plane have negative surface potentials. This is exactly opposite to the KPFM potential map shown in figure 2c (and line scan in figure 3a) and to the integrated CGM profile shown in figure 3c. Correspondingly, the calculated potential gradient (figure 5b and c) is negative moving from an "up" polarised state to a "down" one, whereas the corresponding measured CGM current (figure 1e) is positive. For the head-to-head domain wall region in the x-cut lithium niobate, the calculated potential (figure 5d and f) is a triangular function that is maximised at the wall, whereas in integrated CGM and in directly measured KPFM measurements (figure 3b and d), the triangular function is inverted (with a minimum at the wall). The associated CGM currents are also opposite in sign to those of the calculated potential gradients (compare figure 1d and f to figure 5e and f). In short, while the CGM and KPFM information is largely self-consistent, all information is opposite in sign to that expected from modelling.

We had previously considered it likely that a factor of "-1" could be generated in the CGM data because of the internal circuitry of the scanning probe microscope, effectively monitoring currents from Earth to the tip, rather than the other way around [26] (the same was also seen by Hong *et al.* [23]). However, the KPFM data should not also suffer from this artefact. Indeed, we have tested the KPFM in our scanning probe microscope, by monitoring the potential above a thin film electrode connected to a voltage source, and it was found to be entirely accurate. We must therefore conclude that the measured surface potential (spatially integrated CGM and KPFM) is genuinely opposite in sign to that expected in the absence of screening. In fact, this is commonly observed in the published literature [27-30] and in ambient imaging should, perhaps, be fully expected.

## 3. Conclusion

By using scanning probe microscopy to examine different ferroelectric domain patterns, we present evidence that currents measured by Charge Gradient Microscopy (CGM) are generated by gradients in electrostatic surface potential, rather than gradients in surface bound charge or screening charge densities. This is best illustrated when the spatial integration of the CGM current is compared to the surface potential measured directly by Kelvin Probe Force Microscopy (KPFM). This conclusion makes the interpretation of CGM information relatively straightforward, as it simply offers an alternative to KPFM for determining the surface potential in materials. A possible advantage of the technique is that the strong tip-sample contact needed for CGM measurements means that surface potentials can be accessed even in environments where rapid accumulation of screening adsorbates would normally occur which would obscure KPFM contrast – the CGM scanning continually scrapes most of these adsorbates away.

## 4. Experimental Methods

**Charge Gradient Microscopy:**

Charge gradient microscopy experiments were performed using an Asylum MFP-3D Infinity AFM system in the current-sensing mode (ORCA). The technique involves rastering a grounded conductive probe (25Pt300B from Rocky Mountain Nanotechnology) across the surface of a grounded sample at high scan speeds (~0.5mm/s) with a high deflection setpoint or tip pressure (18V, ~2µN force), while passively measuring any current that flows. The sign of the current flow depends on the scanning direction of the probe and reflects local changes in the electrostatic potential on the surface of the sample. Offset currents induced by the experimental set-up were removed for each sample.

**Kelvin Probe Force Microscopy:**

KPFM experiments were performed using the same AFM system. In these studies, a Pt/Ir-coated Si probe with a resonance frequency of ~70 kHz (Nanosensors, PPP-EFM) was used. KPFM is a two-pass technique, in which the surface topography is mapped during the first pass via conventional tapping mode, before withdrawing to a fixed height above the surface (20nm for our experiments) for the second pass. During this second pass, a d.c bias is applied to the probe such that it matches the contact potential difference between the probe and sample surface. This is done by nullifying the vibration of the probe, which is initially driven by the electrostatic force induced on the AFM probe, giving a true measurement of the local surface potential.

**Piezoresponse Force Microscopy:**

Piezoresponse Force Microscopy experiments were performed with solid Pt probes (25Pt300B from Rocky Mountain Nanotechnology) using the MFP's internal lock-in amplifier. Imaging was carried out near resonance, with a frequency of ~100kHz for observing out of plane domains and ~200kHz for in plane domains. AC biases between 1-5V were used for these measurements.

Validation of the domain structure was carried out using a Veeco Dimension 3100 AFM system (equipped with Nanoscope IIIa controller) in conjunction with an EG&G 7265 lock-in amplifier. For these experiments, Pt/Ir-coated Si probes (Nanosensors, PPP-EFM) were used and measurements were undertaken away from resonance at 20kHz with applied AC biases between 1-5V. Rotating the crystal orientation allowed for the relative polarisation orientations to be unequivocally determined using vector PFM (combination of vertical and lateral PFM).

**Modelling of electrostatic potential from finite element modelling**

Comsol Multiphysics 5.4 was used to model the static electrostatic potential at the surface of dielectric materials with a spontaneous polarisation. The exterior boundary conditions are consistent with what would happen in vacuum, while the interior boundaries are set for the continuity of the dielectric displacements. The mesh used consisted of tetrahedra with the largest dimension smaller than 1 µm and routinely one to two orders of magnitude lower. Simulations were stopped when convergence lower than 0.001 was obtained. The geometry was set to reproduce fairly the experimental geometry: a parallelepiped with width 10 µm, thickness 500 µm and length between 20-40 µm depending on the domain patterns. The thickness was chosen to be similar to the experimental thickness, the width is irrelevant because of the symmetry of the experiment and the length of each domains was chosen to match experiments.

## Acknowledgements

The authors acknowledge funding to support the research presented from the Engineering and Physical Sciences Research Council (EPSRC) in the UK and from the Department for Employment (DfE) in Northern Ireland. RGPMcQ also acknowledges support from a UKRI Future Leaders Fellowship (MR/T043172/1).

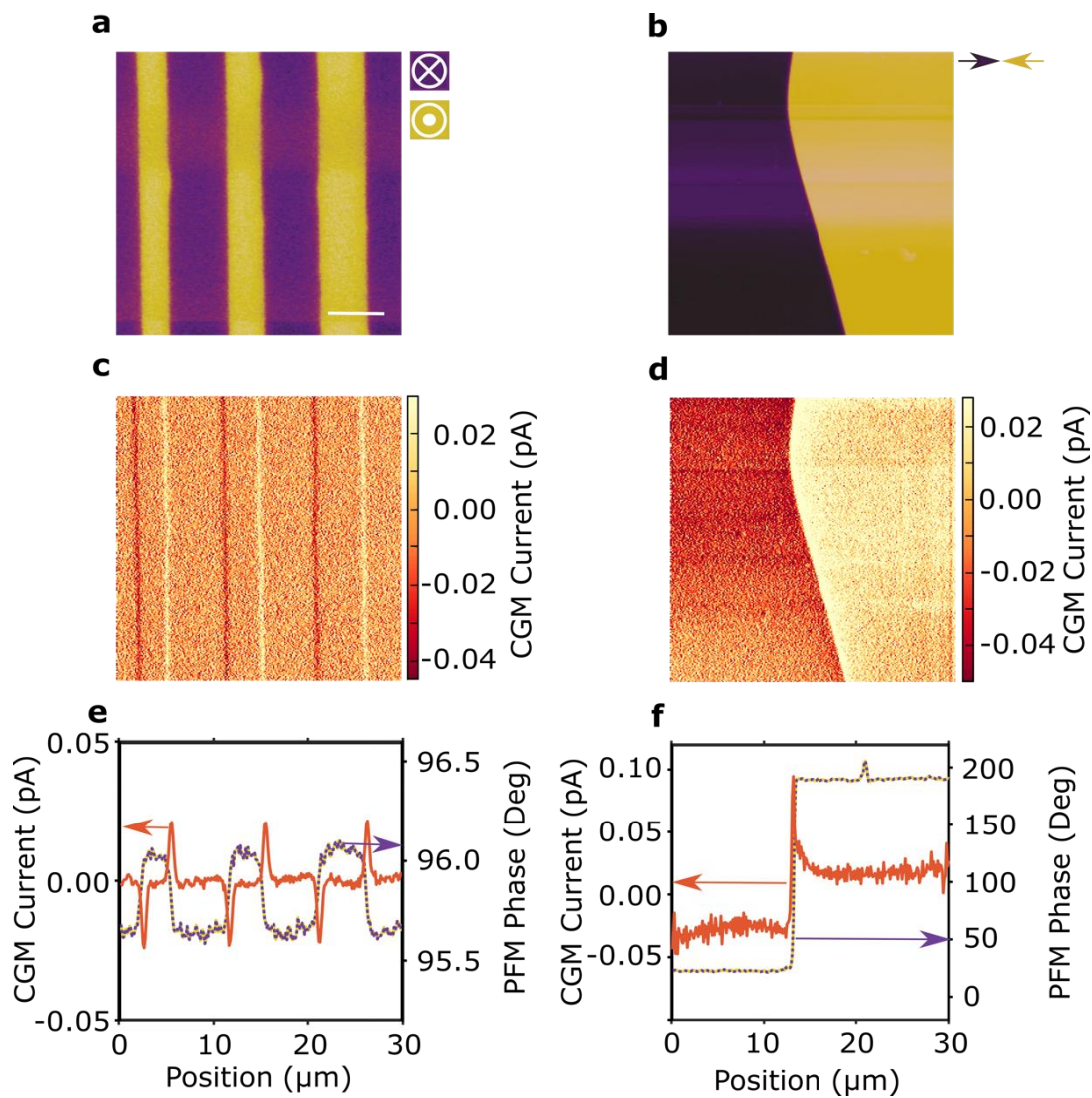

**Figure 1: Charge Gradient Microscopy (CGM) of domains in lithium niobate (LNO).** Piezoresponse force microscopy (PFM) phase maps of (a) c+/c- domains, with polarisation approximately perpendicular to the surface, in z-cut periodically poled LNO (vertical PFM) and (b) head-to-head charged domain wall region, with polarisation approximately parallel to the surface, in x-cut LNO (lateral PFM). Corresponding CGM map for (c) c+/c- domains and (d) head-to-head charged domain wall region. All images are shown at the same magnification and the scale bar in (a) is 6µm long. Line trace information (e, f), taken approximately perpendicular to the surface traces of the domain walls, showing CGM data (orange), summed over multiple scan lines with the slow scan axis disabled, along with PFM phase variations (to help identify the positions of domain walls).

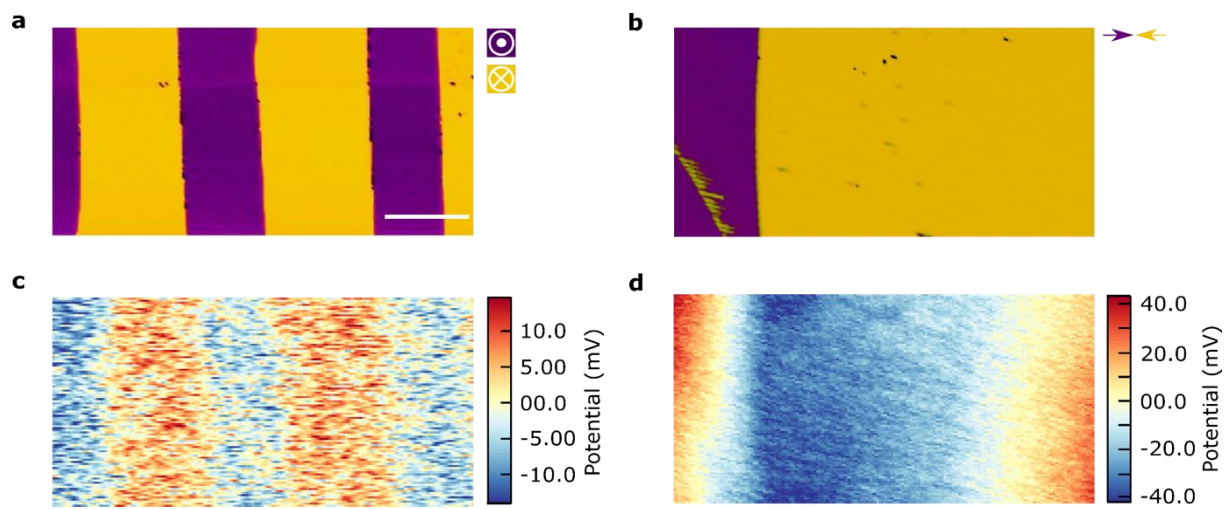

**Figure 2: Kelvin Probe Force Microscopy (KPFM) of the lithium niobate domains**. PFM phase maps for (a) c+/c- domains, with polarisation approximately perpendicular to the surface, in z-cut periodically poled LNO (vertical PFM) and (b) head-to-head charged domain wall region, with polarisation approximately parallel to the surface, in x-cut LNO (lateral PFM), along with associated surface potential maps, measured using KPFM (c, d). Images are all at the same magnification and the scale bar (in (a)) is 4.7µm long.

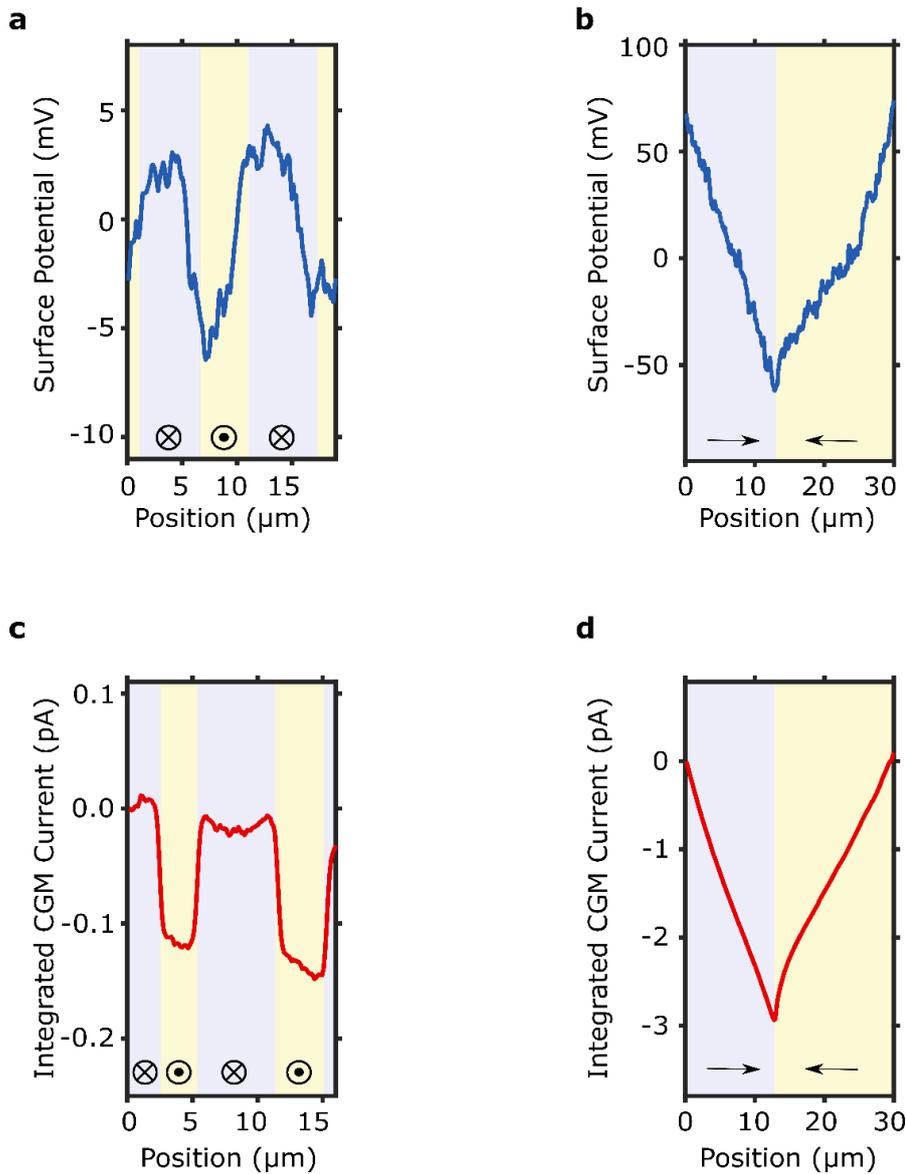

**Figure 3: Line traces comparing the measured surface potential with the spatial integration of the CGM current.** KPFM surface potential variations associated with (a) c+/c- domains in z-cut periodically poled LNO and (b) the head-to-head charged domain wall region in x-cut LNO, along with corresponding plots generated by integrating the CGM currents with respect to distance along the scan direction (c and d). Similarities between the measured potential and the integrated CGM current are obvious. Background colours represent different domain orientations with local polarisation directions as indicated by arrows. Data presented have been obtained from scans in which the slow scan axis has been disabled, such that information is continuously collected for a single scan line and averaged.

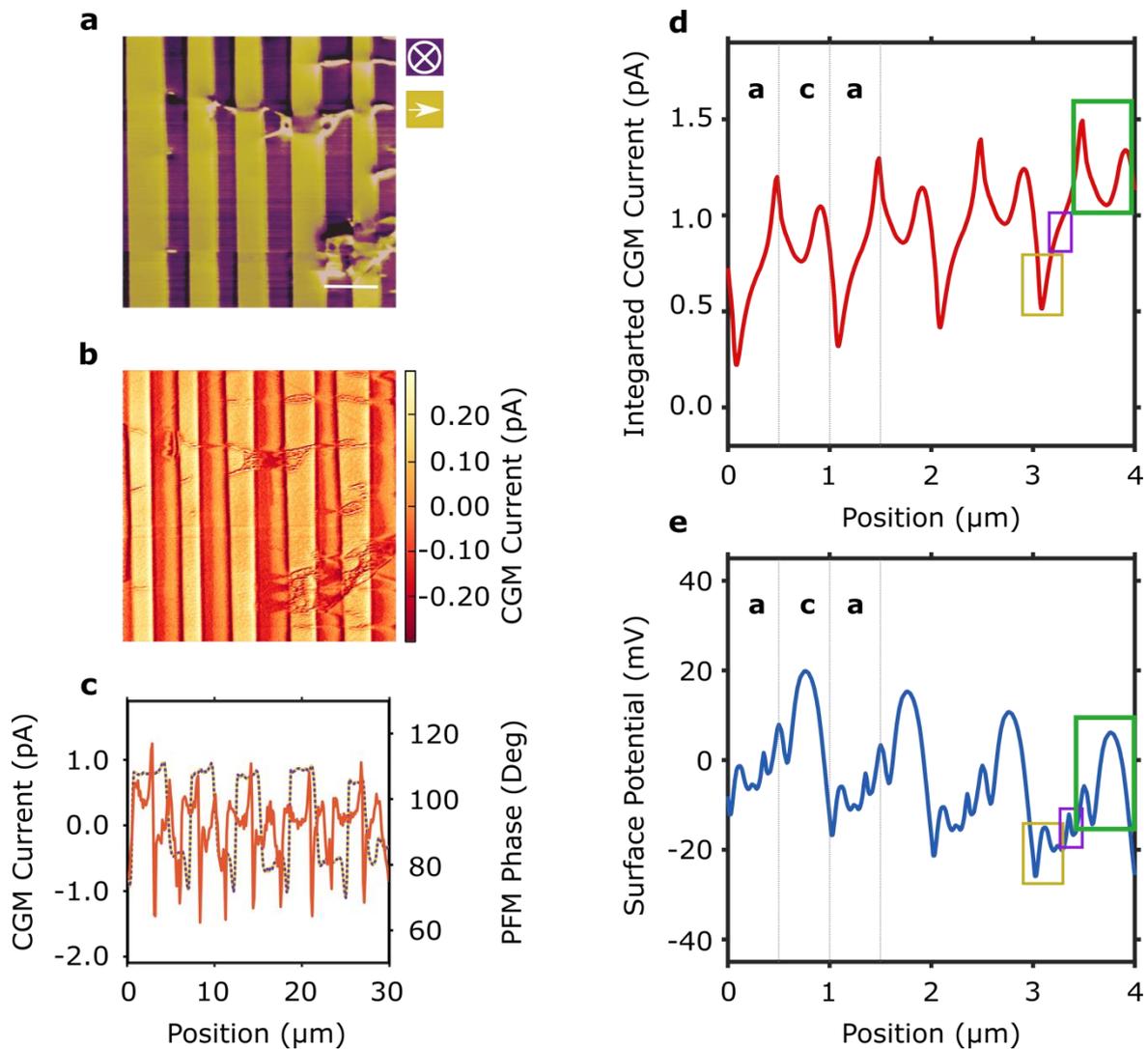

**Figure 4: Relationship between the CGM current and surface potential for a more complex set of domain variants (a/c domains in barium titanate).** (a) PFM phase image (scale bar also applies to (b) and measures 6μm). (b) CGM current map and (c) the corresponding CGM line trace (data from a line scan with the slow scan axis disabled) with PFM phase information overlayed. Data from each a and c domain region was averaged and then the functions stitched together to produce and idealised version of (d) the integrated CGM current and (e) the measured surface potential. While these functions clearly differ, comparisons of the forms of the functions within specific microstructural subregions (identified by the highlighted coloured boxes) suggest that they might both be imperfectly measuring the same thing.

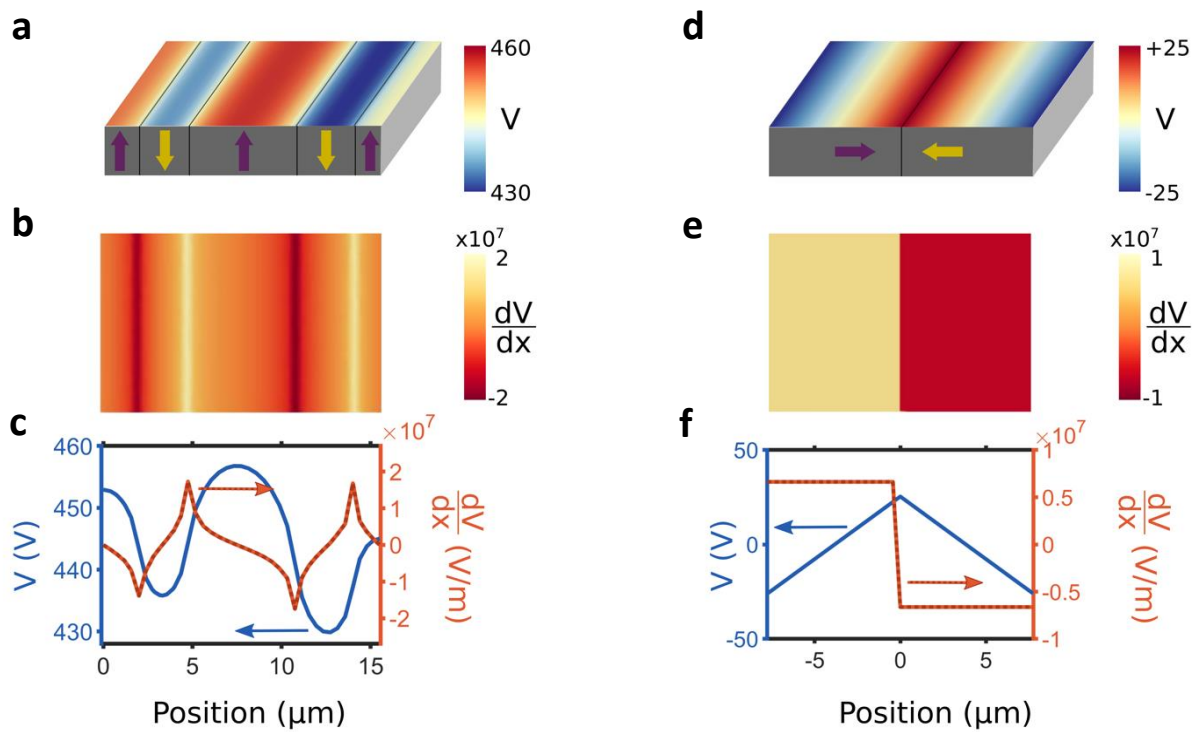

**Figure 5: Finite element simulation of the surface potential and associated gradients for LNO.** (a,d) colour maps of the surface potential *V*, (b,e) colour maps of the positional derivative of the surface potential $\frac{dV}{dx}$ and (c,f) comparison of the line profiles along *x* of the surface potential *V* (blue) and spatial derivative $\frac{dV}{dx}$ (dotted orange). The spatial derivatives of the potential scale with the measured CGM but are inverted in sign.